\pdfoutput=1

\documentclass[english,a4paper,12pt]{article}
\usepackage{varioref}
\usepackage{latexsym}
\usepackage{amsmath}
\usepackage{tabularx}

\numberwithin{equation}{section}
\input{tcilatex}
\usepackage{graphicx}

\begin{document}

\title{The equations of the ideal latches}
\author{Serban E. Vlad \\
The computers department, \\
Oradea City Hall, Oradea, Romania\\
web: www.geocities.com/serban\_e\_vlad}
\date{}
\maketitle

\begin{center}
\begin{tabular}{p{12cm}}
\hline
\textbf{Abstract}. {\footnotesize The latches are simple circuits with
feedback from the digital electrical engineering. We have included in our
work the C element of Muller, the RS latch, the clocked RS latch, the D
latch and also circuits containing two interconnected latches: the edge
triggered RS flip-flop, the D flip-flop, the JK flip-flop, the T flip-flop.
The purpose of this study is to model with equations the previous circuits,
considered to be ideal, i.e. non-inertial. The technique of analysis is the
pseudoboolean differential calculus.} \hline
\end{tabular}
\end{center}

\section{Latches, the general equation}

$\mathbf{B}=\{0,1\}$ is the Boole algebra with two elements. The (normal)
signals are by definition the functions $x:\mathbf{R}\longrightarrow \mathbf{%
B}$ of the form 
\begin{equation*}
x(t)=x(\tau _{0}-0)\cdot \varphi _{(-\infty ,\tau _{0})}(t)\oplus x(\tau
_{0})\cdot \varphi _{\lbrack \tau _{0},\tau _{1})}(t)\oplus x(\tau
_{1})\cdot \varphi _{\lbrack \tau _{1},\tau _{2})}(t)\oplus ...
\end{equation*}%
where $\mathbf{R}$ is the time set, $\varphi _{()}:\mathbf{R}\rightarrow 
\mathbf{B}$ is the characteristic function and $0\leq \tau _{0}<\tau
_{1}<\tau _{2}<...$ is an unbounded sequence. The equations of the (ideal)
latches consist in the next system%
\begin{equation}
\left\{ 
\begin{array}{c}
\overline{x(t-0)}\cdot x(t)=\overline{x(t-0)}\cdot u(t) \\ 
x(t-0)\cdot \overline{x(t)}=x(t-0)\cdot v(t) \\ 
u(t)\cdot v(t)=0%
\end{array}%
\right.
\end{equation}%
where $u,v,x$ are signals and $x$ is the unknown. The last equation of the
system is called the admissibility condition (of the inputs). In order to
solve the system (1.1) we associate to the functions $u,v$ the next sets $%
U_{2k},V_{2k+1}$ and respectively numbers $t_{k}$:%
\begin{eqnarray*}
U_{0} &=&\{t|\overline{u(t-0)}\cdot u(t)=1\},\quad \quad \quad \quad \quad
t_{0}=\min U_{0} \\
V_{1} &=&\{t|\overline{v(t-0)}\cdot v(t)=1,t>t_{0}\},\quad \quad t_{1}=\min
V_{1} \\
U_{2} &=&\{t|\overline{u(t-0)}\cdot u(t)=1,t>t_{1}\},\quad \quad t_{2}=\min
U_{2} \\
V_{3} &=&\{t|\overline{v(t-0)}\cdot v(t)=1,t>t_{2}\},\quad \quad t_{3}=\min
V_{3} \\
&&...
\end{eqnarray*}%
and the next inclusions, respectively inequalities are true:%
\begin{equation*}
U_{0}\supset U_{2}\supset U_{4}\supset ...\quad V_{1}\supset V_{3}\supset
V_{5}\supset ...\quad
\end{equation*}%
\begin{equation*}
0\leq t_{0}<t_{1}<t_{2}<...
\end{equation*}%
For each of $U_{2k}$ ($V_{2k+1}$) we have the possibilities:

- it is empty. Then $t_{2k}$ ($t_{2k+1}$) is undefined and all $%
U_{2k},V_{2k+1},t_{k}$ of higher rank are undefined

- it is non-empty, finite or infinite. $t_{2k}$ ($t_{2k+1}$) is defined

If $U_{2k}$ ($V_{2k+1}$) are defined for all $k\in \mathbf{N}$, then the
sequence $(t_{k})$ is unbounded.

A similar discussion is related with the sets $V_{2k}^{\prime
},U_{2k+1}^{\prime }$ and respectively numbers $t_{k}^{\prime }:$%
\begin{eqnarray*}
V_{0}^{\prime } &=&\{t|\overline{v(t-0)}\cdot v(t)=1\},\quad \quad \quad
\quad \quad t_{0}^{\prime }=\min V_{0}^{\prime } \\
U_{1}^{\prime } &=&\{t|\overline{u(t-0)}\cdot u(t)=1,t>t_{0}^{\prime
}\},\quad \quad t_{1}^{\prime }=\min U_{1}^{\prime } \\
V_{2}^{\prime } &=&\{t|\overline{v(t-0)}\cdot v(t)=1,t>t_{1}^{\prime
}\},\quad \quad t_{2}^{\prime }=\min V_{2}^{\prime } \\
U_{3}^{\prime } &=&\{t|\overline{u(t-0)}\cdot u(t)=1,t>t_{2}^{\prime
}\},\quad \quad t_{3}^{\prime }=\min U_{3}^{\prime } \\
&&...
\end{eqnarray*}%
For solving the system (1.1) we observe that the unbounded sequence $0\leq
t_{0}^{"}<t_{1}^{"}<t_{2}^{"}<...$ exists with the property that $u,v,x$ are
constant in each of the intervals $(-\infty
,t_{0}^{"}),[t_{0}^{"},t_{1}^{"}),[t_{1}^{"},t_{2}^{"}),...$ where the first
two equations of (1.1) take one of the forms%
\begin{equation}
\left\{ 
\begin{array}{c}
\overline{x(t-0)}\cdot x(t)=\overline{x(t-0)} \\ 
x(t-0)\cdot \overline{x(t)}=0%
\end{array}%
\right.
\end{equation}%
\begin{equation}
\left\{ 
\begin{array}{c}
\overline{x(t-0)}\cdot x(t)=0 \\ 
x(t-0)\cdot \overline{x(t)}=x(t-0)%
\end{array}%
\right.
\end{equation}%
\begin{equation}
\left\{ 
\begin{array}{c}
\overline{x(t-0)}\cdot x(t)=0 \\ 
x(t-0)\cdot \overline{x(t)}=0%
\end{array}%
\right.
\end{equation}%
as $u(t),v(t)$ are equal with $1,0;0,1;0,0$ in those intervals. The
solutions were written in the next table%
\begin{equation*}
\underset{Table\quad 1}{\frame{$%
\begin{array}{cccc}
& 
\begin{array}{c}
eq\quad (1.2) \\ 
u(t)=1,v(t)=0%
\end{array}
& 
\begin{array}{c}
eq\quad (1.3) \\ 
u(t)=0,v(t)=1%
\end{array}
& 
\begin{array}{c}
eq\quad (1.4) \\ 
u(t)=v(t)=0%
\end{array}
\\ 
t\in (-\infty ,t_{0}^{"}) & x(t)=1 & x(t)=0 & 
\begin{array}{c}
x(t)=0 \\ 
x(t)=1%
\end{array}
\\ 
t\in \lbrack t_{k}^{"},t_{k+1}^{"}) & x(t)=1 & x(t)=0 & x(t)=x(t_{k}^{"}-0)%
\end{array}%
$}}
\end{equation*}%
\textbf{Theorem} Equation (1.1) is equivalent with the equation%
\begin{equation}
x(t)\cdot u(t)\cdot \overline{v(t)}\cup \overline{x(t)}\cdot \overline{u(t)}%
\cdot v(t)\cup
\end{equation}%
\begin{equation*}
\cup (\overline{x(t-0)}\cdot \overline{x(t)}\cup x(t-0)\cdot x(t))\cdot 
\overline{u(t)}\cdot \overline{v(t)}=1
\end{equation*}%
\textbf{Proof} The proof is elementary and it is omitted.

Equation (1.5) contains three exclusive possibilities: $x(t)\cdot u(t)\cdot 
\overline{v(t)}=1,$ $\overline{x(t)}\cdot \overline{u(t)}\cdot v(t)=1,$
respectively $(\overline{x(t-0)}\cdot \overline{x(t)}\cup x(t-0)\cdot
x(t))\cdot \overline{u(t)}\cdot \overline{v(t)}=1$ equivalent with (1.2),
(1.3), (1.4).

We solve the system (1.1).

\emph{Case a)} $u(0-0)=0,v(0-0)=0$

$x(0-0)=0$ and $x(0-0)=1$ are both possible. In order to make a distinction
between the two solutions of (1.1) corresponding to the initial value $0$,
respectively to the initial value $1$ we shall note them with $x$,
respectively with $x^{\prime }$.

a.i) $x(0-0)=0$

a.i.1) $U_{0}=\emptyset $

the solution of (1.1) is $x(t)=0$

a.i.2) $U_{0}\neq \emptyset $

and $\exists \varepsilon >0,x(t)=\varphi _{\lbrack t_{0},\infty )}(t)$ for $%
t<t_{0}+\varepsilon $. This fact results by solving (1.4) for $t<t_{0}$ and
then (1.2) followed perhaps by a finite sequence of (1.4), (1.2),
(1.4),\ldots\ in some interval $[t_{0},t_{0}+\varepsilon )$. Furthermore

a.i.2.1) $V_{1}=\emptyset $

the solution of (1.1) is $x(t)=\varphi _{\lbrack t_{0},\infty )}(t)$.

a.i.2.2) $V_{1}\neq \emptyset $

and $\exists \varepsilon >0,x(t)=\varphi _{\lbrack t_{0},t_{1})}(t)$ for $%
t<t_{1}+\varepsilon $. In some interval $[t_{1},t_{1}+\varepsilon )$, we
solved (1.3) followed perhaps by a finite sequence of (1.4), (1.3),
(1.4),\ldots

a.i.2.2.1) $U_{2}=\emptyset $

the solution of (1.1) is $x(t)=\varphi _{\lbrack t_{0},t_{1})}(t)$

a.i.2.2.2) $U_{2}\neq \emptyset $

and $\exists \varepsilon >0,x(t)=\varphi _{\lbrack t_{0},t_{1})}(t)\oplus
\varphi _{\lbrack t_{2},\infty )}(t)$ for $t<t_{2}+\varepsilon $.

a.i.2.2.2.1) $V_{3}=\emptyset $

the solution of (1.1) is $x(t)=\varphi _{\lbrack t_{0},t_{1})}(t)\oplus
\varphi _{\lbrack t_{2},\infty )}(t)$

a.i.2.2.2.2) $V_{3}\neq \emptyset $

...

a.ii) $x^{\prime }(0-0)=1$

a.ii.1) $V_{0}^{\prime }=\emptyset $

the solution of (1.1) is $x^{\prime }(t)=1$

a.ii.2) $V_{0}^{\prime }\neq \emptyset $

$\exists \varepsilon >0,x^{\prime }(t)=\varphi _{(-\infty ,t_{0}^{\prime
})}(t)$ for all $t<t_{0}^{\prime }+\varepsilon $

a.ii.2.1) $U_{1}^{\prime }=\emptyset $

the solution of (1.1) is $x^{\prime }(t)=\varphi _{(-\infty ,t_{0}^{\prime
})}(t)$

a.ii.2.2) $U_{1}^{\prime }\neq \emptyset $

$\exists \varepsilon >0,x^{\prime }(t)=\varphi _{(-\infty ,t_{0}^{\prime
})}(t)\oplus \varphi _{\lbrack t_{1}^{\prime },\infty )}(t)$ for all $%
t<t_{1}^{\prime }+\varepsilon $

...

We have drawn in Figures 1 and 2 the solutions $x,x^{\prime }$ corresponding
to Case a) in the situation when $t_{0}<t_{0}^{\prime },$ respectively when $%
t_{0}>t_{0}^{\prime }$ (the equality $t_{0}=t_{0}^{\prime }$ is impossible,
because it implies $u(t_{0})=v(t_{0}^{\prime })=1$, contradiction with
(1.1)). We observe the fact that $x_{|[t_{0},\infty )}=x_{|[t_{0},\infty
)}^{\prime }$, respectively $x_{|[t_{0}^{\prime },\infty
)}=x_{|[t_{0}^{\prime },\infty )}^{\prime }$ thus after the first common
value of the (distinct) solutions $x,x^{\prime }$ they coincide.\FRAME{%
ftbpFUX}{5.1076in}{3.0753in}{0pt}{\Qcb{Case a), $t_{0}<t_{0}^{\prime }$}}{}{%
f1.jpg}{\raisebox{-3.0753in}{\includegraphics[height=3.0753in]{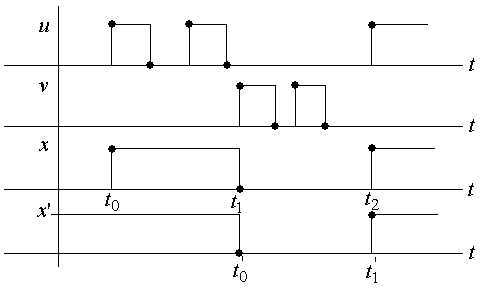}}}\FRAME{ftbpFUX}{5.0548in}{3.1168in}{0pt}{%
\Qcb{Case a), $t_{0}>t_{0}^{\prime }$}}{}{f2.jpg}{\raisebox{-3.1168in}{\includegraphics[height=3.1168in]{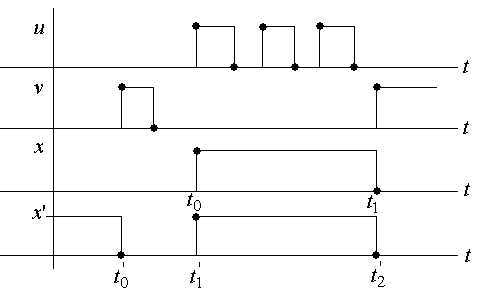}}}

\emph{Case b)} $u(0-0)=1,v(0-0)=0$

the only possibility is $x(0-0)=1$

b.1) $V_{0}^{\prime }=\emptyset $

the solution of (1.1) is $x(t)=1$

b.2) $V_{0}^{\prime }\neq \emptyset $

$\exists \varepsilon >0,x(t)=\varphi _{(-\infty ,t_{0}^{\prime })}(t)$ for
all $t<t_{0}^{\prime }+\varepsilon $

...

\emph{Case c)} $u(0-0)=0,v(0-0)=1$

the only possibility is $x(0-0)=0$

c.1) $U_{0}=\emptyset $

the solution of (1.1) is $x(t)=0$

c.2) $U_{0}\neq \emptyset $

$\exists \varepsilon >0,x(t)=\varphi _{\lbrack t_{0},\infty )}(t)$ for $%
t<t_{0}+\varepsilon $

...

We have proved the next

\textbf{Theorem} If $u(t)=v(t)=0$, the system (1.1) has two solutions $%
x(t)=0 $ and $x(t)=1$. If $u(0-0)=v(0-0)=0$ but $\exists t>0,u(t)\cup v(t)=1$%
, then (1.1) has two distinct solutions corresponding to $x(0-0)=0$ and $%
x(0-0)=1$, that become equal at the first time instant $t>0$ when $u(t)\cup
v(t)=1$. And if $u(0-0)\cup v(0-0)=1$, then the solution is unique.

\section{C element}

We call the equations of the C element of Muller any of the next equivalent
statements:%
\begin{equation}
\left\{ 
\begin{array}{c}
\overline{x(t-0)}\cdot x(t)=\overline{x(t-0)}\cdot u(t)\cdot v(t) \\ 
x(t-0)\cdot \overline{x(t)}=x(t-0)\cdot \overline{u(t)}\cdot \overline{v(t)}%
\end{array}%
\right.
\end{equation}%
and respectively\FRAME{ftbpFUX}{2.6039in}{2.6143in}{0pt}{\Qcb{The C element
of Muller}}{}{c_element1.jpg}{\raisebox{-2.6143in}{\includegraphics[height=2.6143in]{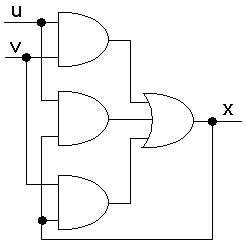}}}\FRAME{ftbpFUX}{1.7876in}{1.0222in}{%
0pt}{\Qcb{The symbol of the C element of Muller}}{}{c_element_symbol1.jpg}{%
\raisebox{-1.0222in}{\includegraphics[height=1.0222in]{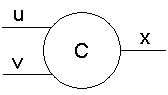}}}%
\begin{equation}
x(t)\cdot u(t)\cdot v(t)\cup \overline{x(t)}\cdot \overline{u(t)}\cdot 
\overline{v(t)}\cup
\end{equation}%
\begin{equation}
\cup (\overline{x(t-0)}\cdot \overline{x(t)}\cup x(t-0)\cdot x(t))\cdot (%
\overline{u(t)}\cdot v(t)\cup u(t)\cdot \overline{v(t)})=1  \notag
\end{equation}%
where $u,v,x$ are signals, the first two called inputs and the last --
state. Equations (2.1), (2.2) are the equations of a latch (1.1), (1.5)
where $u(t)$ is replaced by $u(t)\cdot v(t)$ and $v(t)$ is replaced by $%
\overline{u(t)}\cdot \overline{v(t)}$. It is observed the satisfaction of
the admissibility condition of the inputs. The analysis of (2.2) is obvious: 
$x(t)$ is $1$ if $u(t)=v(t)=1$, $x(t)$ is $0$ if $u(t)=v(t)=0$ and $%
x(t)=x(t-0),\quad x(t)$ keeps its previous value otherwise. The general form
of equations (2.1), (2.2) for $m$ inputs $u_{1},...,u_{m}$ is%
\begin{equation*}
\left\{ 
\begin{array}{c}
\overline{x(t-0)}\cdot x(t)=\overline{x(t-0)}\cdot u_{1}(t)\cdot ...\cdot
u_{m}(t) \\ 
x(t-0)\cdot \overline{x(t)}=x(t-0)\cdot \overline{u_{1}(t)}\cdot ...\cdot 
\overline{u_{m}(t)}%
\end{array}%
\right.
\end{equation*}%
\begin{equation*}
x(t)\cdot u_{1}(t)\cdot ...\cdot u_{m}(t)\cup \overline{x(t)}\cdot \overline{%
u_{1}(t)}\cdot ...\cdot \overline{u_{m}(t)}\cup
\end{equation*}%
\begin{equation}
\cup (\overline{x(t-0)}\cdot \overline{x(t)}\cup x(t-0)\cdot x(t))\cdot 
\overline{u_{1}(t)\cdot ...\cdot u_{m}(t)}\cdot (u_{1}(t)\cup ...\cup
u_{m}(t))=1  \notag
\end{equation}

\section{RS latch}

The equations of the RS latch are given by%
\begin{equation}
\left\{ 
\begin{array}{c}
\overline{Q(t-0)}\cdot Q(t)=\overline{Q(t-0)}\cdot S(t) \\ 
Q(t-0)\cdot \overline{Q(t)}=Q(t-0)\cdot R(t) \\ 
R(t)\cdot S(t)=0%
\end{array}%
\right. 
\end{equation}%
and equivalently by%
\begin{equation}
Q(t)\cdot \overline{R(t)}\cdot S(t)\cup \overline{Q(t)}\cdot R(t)\cdot 
\overline{S(t)}\cup 
\end{equation}%
\begin{equation*}
\cup (\overline{Q(t-0)}\cdot \overline{Q(t)}\cup Q(t-0)\cdot Q(t))\cdot 
\overline{R(t)}\cdot \overline{S(t)}=1
\end{equation*}%
In (3.1), (3.2) $R,S,Q$ are signals. $R,S$ are called inputs: the reset
input and the set input and $Q$ is the state, the unknown relative to which
the equations are solved. These equations coincide with (1.1) and (1.5) but
the notations are different and traditional.\FRAME{ftbpFUX}{1.6717in}{%
1.8187in}{0pt}{\Qcb{The RS latch circuit}}{}{rs_latch1.jpg}{\raisebox{-1.8187in}%
{\includegraphics[height=1.8187in]{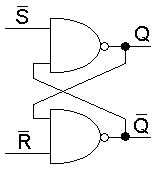}}}
\FRAME{ftbpFUX}{1.2747in}{0.9288in}{0pt}{\Qcb{The symbol of the
RS latch}}{}{f4b.jpg}{\raisebox{-0.9288in}{\includegraphics[height=0.9288in]{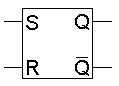}}} We conclude the things that were
discussed in section 1 by the next statements related with equation (3.2).
At the RS latch, $Q(t)=1$ if $R(t)=0,S(t)=1;$ $Q(t)=0$ if $R(t)=1,S(t)=0;$
and $Q(t)=Q(t-0),Q$ keeps its previous value if $R(t)=0,S(t)=0.$

\section{Clocked RS latch}

The equivalent statements%
\begin{equation}
\left\{ 
\begin{array}{c}
\overline{Q(t-0)}\cdot Q(t)=\overline{Q(t-0)}\cdot S(t)\cdot C(t) \\ 
Q(t-0)\cdot \overline{Q(t)}=Q(t-0)\cdot R(t)\cdot C(t) \\ 
R(t)\cdot S(t)\cdot C(t)=0%
\end{array}%
\right.
\end{equation}%
and%
\begin{equation*}
C(t)\cdot (Q(t)\cdot \overline{R(t)}\cdot S(t)\cup \overline{Q(t)}\cdot
R(t)\cdot \overline{S(t)}\cup
\end{equation*}%
\begin{equation}
\cup (\overline{Q(t-0)}\cdot \overline{Q(t)}\cup Q(t-0)\cdot Q(t))\cdot 
\overline{R(t)}\cdot \overline{S(t)})\cup
\end{equation}%
\begin{equation*}
\cup \overline{C(t)}\cdot (\overline{Q(t-0)}\cdot \overline{Q(t)}\cup
Q(t-0)\cdot Q(t))=1
\end{equation*}%
are called the equations of the clocked RS latch. $R,S,C,Q$ are signals:.
the reset, the set and the clock input, respectively the state. The
equations (4.1), (4.2) result from (1.1) and (1.5) where $u(t)=S(t)\cdot
C(t),$ $v(t)=R(t)\cdot C(t)$.\FRAME{ftbpFUX}{2.866in}{2.143in}{0pt}{\Qcb{The
clocked RS latch circuit}}{}{clocked_rs_latch1.jpg}{\raisebox{-2.143in}{\includegraphics[height=2.143in]{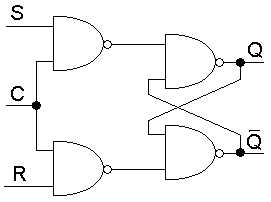}}} \FRAME{ftbpFUX}{1.2531in}{0.8761in}{%
0pt}{\Qcb{The symbol of the clocked RS latch}}{}{f5b.jpg}{\raisebox{-0.8761in}{\includegraphics[height=0.8761in]{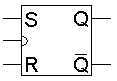}}%
}The clocked RS latch behaves like an RS latch when $C(t)=1$ and keeps the
state constant $Q(t)=Q(t-0)$ when $C(t)=0.$

\section{D latch}

We call the equations of the D latch any of the next equivalent statements%
\begin{equation}
\left\{ 
\begin{array}{c}
\overline{Q(t-0)}\cdot Q(t)=\overline{Q(t-0)}\cdot D(t)\cdot C(t) \\ 
Q(t-0)\cdot \overline{Q(t)}=Q(t-0)\cdot \overline{D(t)}\cdot C(t)%
\end{array}%
\right.
\end{equation}%
and respectively%
\begin{equation}
C(t)\cdot (\overline{Q(t)}\cdot \overline{D(t)}\cup Q(t)\cdot D(t))\cup 
\overline{C(t)}\cdot (\overline{Q(t-0)}\cdot \overline{Q(t)}\cup Q(t-0)\cdot
Q(t))=1
\end{equation}%
$D,C,Q$ are signals: the data input $D$, the clock input $C$ and the state $%
Q $. On one hand, from (5.1) it is seen the satisfaction of the
admissibility condition of the inputs. And on the other hand (5.1), (5.2)
result from the equations of the clocked RS latch (4.1), (4.2) where $R=%
\overline{S\cdot C}$ and we have used the traditional notation $D$ for the
data input, instead of $S$.\FRAME{ftbpFUX}{2.8867in}{2.1638in}{0pt}{\Qcb{The
D latch circuit}}{}{d_latch1.jpg}{\raisebox{-2.1638in}{\includegraphics[height=2.1638in]{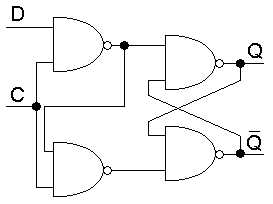}}} \FRAME{ftbpFUX}{1.3474in}{0.9392in}{%
0pt}{\Qcb{The symbol of the D latch}}{}{f6b.jpg}{\raisebox{-0.9392in}{\includegraphics[height=0.9392in]{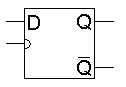}}%
}When $C(t)=1,$ the D latch makes $Q(t)=D(t)$ and when $C(t)=0,$ $Q$ is
constant.

\section{Edge triggered RS flip-flop}

Any of the equivalent statements%
\begin{equation}
\left\{ 
\begin{array}{c}
\overline{P(t-0)}\cdot P(t)=\overline{P(t-0)}\cdot S(t)\cdot C(t) \\ 
P(t-0)\cdot \overline{P(t)}=P(t-0)\cdot R(t)\cdot C(t) \\ 
R(t)\cdot S(t)\cdot C(t)=1 \\ 
\overline{Q(t-0)}\cdot Q(t)=\overline{Q(t-0)}\cdot P(t)\cdot \overline{C(t)}
\\ 
Q(t-0)\cdot \overline{Q(t)}=Q(t-0)\cdot \overline{P(t)}\cdot \overline{C(t)}%
\end{array}%
\right.
\end{equation}%
and respectively\FRAME{ftbpFUX}{4.8456in}{2.5624in}{0pt}{\Qcb{The edge
triggered RS flip-flop circuit}}{}{edge_triggered1.jpg}{\raisebox{-2.5624in}{\includegraphics[height=2.5624in]{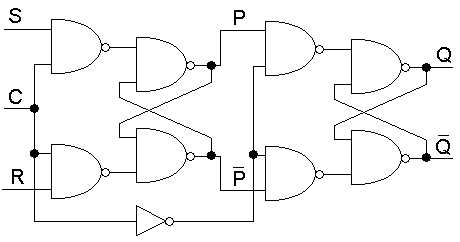}}}\FRAME{ftbpFUX}{1.2315in}{0.8864in%
}{0pt}{\Qcb{The symbol of the edge triggered RS flip-flop}}{}{f7b.jpg}{%
\raisebox{-0.8864in}{\includegraphics[height=0.8864in]{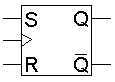}}%
}%
\begin{equation*}
C(t)\cdot (\overline{Q(t-0)}\cdot \overline{Q(t)}\cup Q(t-0)\cdot Q(t))\cdot
(P(t)\cdot \overline{R(t)}\cdot S(t)\cup
\end{equation*}%
\begin{equation}
\cup \overline{P(t)}\cdot R(t)\cdot \overline{S(t)}\cup (\overline{P(t-0)}%
\cdot \overline{P(t)}\cup P(t-0)\cdot P(t))\cdot \overline{R(t)}\cdot 
\overline{S(t)})\cup
\end{equation}%
\begin{equation*}
\cup \overline{C(t)}\cdot (\overline{Q(t)}\cdot \overline{P(t-0)}\cdot 
\overline{P(t)}\cup Q(t)\cdot P(t-0)\cdot P(t))=1
\end{equation*}%
is called the equation of the edge triggered RS flip-flop. $R,S,C,P,Q$ are
signals: the reset input $R$, the set input $S$, the clock input $C$, the
next state $P$ and the state $Q$. In (6.1), (6.2) the signals $R,S,C,P$ and $%
P,\overline{C},Q$ satisfy the equations of a clocked RS latch and of a D
latch and (6.2) represents the term by term product of (4.2) with (5.2)
written with these variables. The two latches are called master and slave.
The name of edge triggered RS flip-flop refers to the fact that $Q(t)$ is
constant at all time instances except $C(t-0)\cdot \overline{C(t)}=1,$ when $%
Q(t)=P(t-0)=\left\{ 
\begin{array}{c}
1,if\quad R(t-0)=0,S(t-0)=1 \\ 
0,if\quad R(t-0)=1,S(t-0)=0%
\end{array}%
\right. ,$ this is the so called 'falling edge' of the clock input.

\section{D flip-flop}

We call the equations of the D flip-flop any of the next equivalent
conditions:\FRAME{ftbpFUX}{4.8974in}{2.6039in}{0pt}{\Qcb{The D flip-flop
circuit}}{}{d_flip_flop1.jpg}{\raisebox{-2.6039in}{\includegraphics[height=2.6039in]{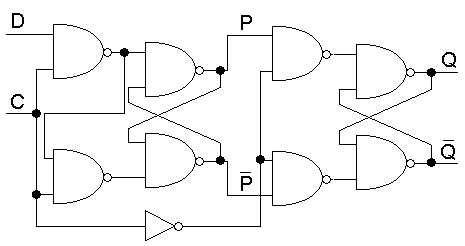}}}\FRAME{ftbpFUX}{1.3266in}{0.9288in%
}{0pt}{\Qcb{The symbol of the D flip-flop}}{}{f8b.jpg}{\raisebox{-0.9288in}{\includegraphics[height=0.9288in]{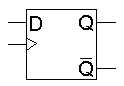}}%
}%
\begin{equation}
\left\{ 
\begin{array}{c}
\overline{P(t-0)}\cdot P(t)=\overline{P(t-0)}\cdot D(t)\cdot C(t) \\ 
P(t-0)\cdot \overline{P(t)}=P(t-0)\cdot \overline{D(t)}\cdot C(t) \\ 
\overline{Q(t-0)}\cdot Q(t)=\overline{Q(t-0)}\cdot P(t)\cdot \overline{C(t)}
\\ 
Q(t-0)\cdot \overline{Q(t)}=Q(t-0)\cdot \overline{P(t)}\cdot \overline{C(t)}%
\end{array}%
\right.
\end{equation}%
and respectively%
\begin{equation}
C(t)\cdot (\overline{Q(t-0)}\cdot \overline{Q(t)}\cup Q(t-0)\cdot Q(t))\cdot
(\overline{P(t)}\cdot \overline{D(t)}\cup P(t)\cdot D(t))\cup
\end{equation}%
\begin{equation*}
\cup \overline{C(t)}\cdot (\overline{Q(t)}\cdot \overline{P(t-0)}\cdot 
\overline{P(t)}\cup Q(t)\cdot P(t-0)\cdot P(t))=1
\end{equation*}%
$D,C,P,Q$ are signals, called: the data input $D$, the clock input $C$, the
next state $P$ and the state $Q$. We observe that the equations of the $D$
flip-flop represent the special case of edge triggered RS flip-flop when $R=%
\overline{S\cdot C}$ and $S$ was noted with $D$. The D flip-flop has the
state $Q$ constant except for the time instants when $C(t-0)\cdot \overline{%
C(t)}=1$; then $Q(t)=D(t-0)$.

\section{JK flip-flop}

The equivalent statements:%
\begin{equation}
\left\{ 
\begin{array}{c}
\overline{P(t-0)}\cdot P(t)=\overline{P(t-0)}\cdot J(t)\cdot \overline{Q(t)}%
\cdot C(t) \\ 
P(t-0)\cdot \overline{P(t)}=P(t-0)\cdot K(t)\cdot Q(t)\cdot C(t) \\ 
\overline{Q(t-0)}\cdot Q(t)=\overline{Q(t-0)}\cdot P(t)\cdot \overline{C(t)}
\\ 
Q(t-0)\cdot \overline{Q(t)}=Q(t-0)\cdot \overline{P(t)}\cdot \overline{C(t)}%
\end{array}%
\right.
\end{equation}%
and%
\begin{equation*}
C(t)\cdot (\overline{Q(t-0)}\cdot \overline{Q(t)}\cup Q(t-0)\cdot Q(t))\cdot
(P(t)\cdot J(t)\cdot \overline{Q(t)}\cup \overline{P(t)}\cdot K(t)\cdot
Q(t)\cup
\end{equation*}%
\begin{equation}
\cup (\overline{P(t-0)}\cdot \overline{P(t)}\cup P(t-0)\cdot P(t))\cdot (%
\overline{J(t)}\cdot \overline{K(t)}\cup \overline{J(t)}\cdot \overline{Q(t)}%
\cup \overline{K(t)}\cdot Q(t)))\cup
\end{equation}%
\begin{equation*}
\cup \overline{C(t)}\cdot (\overline{Q(t)}\cdot \overline{P(t-0)}\cdot 
\overline{P(t)}\cup Q(t)\cdot P(t-0)\cdot P(t))=1
\end{equation*}%
are called the equations of the JK flip-flop. $J,K,C,P,Q$ are signals: the J
input, the K input, the clock input C, the next state P and the state Q. The
first two equations of (8.1) (modeling the master latch) coincide with the
first two equations of the edge triggered RS flip-flop where $S(t)=J(t)\cdot 
\overline{Q(t)},$ $R(t)=K(t)\cdot Q(t)$ and the last two equations of (8.1)
(modeling the slave latch) coincide with the last two equations of the edge
triggered RS flip-flop. We observe that the conditions of admissibility of
the inputs of the master and of the slave latch are fulfilled. To be
compared (8.2) and (6.2). The JK flip-flop is similar with the edge
triggered flip-flop, for example $Q$ changes value only when $C(t-0)\cdot 
\overline{C(t)}=1.$ Let $C(t)=1;$ because $Q(t)=Q(t-0)$ i.e. $Q$ is
constant, in the reunion\FRAME{ftbpFUX}{4.8248in}{2.7614in}{0pt}{\Qcb{The JK
flip-flop circuit}}{}{jk_flip_flop1.jpg}{\raisebox{-2.7614in}{\includegraphics[height=2.7614in]{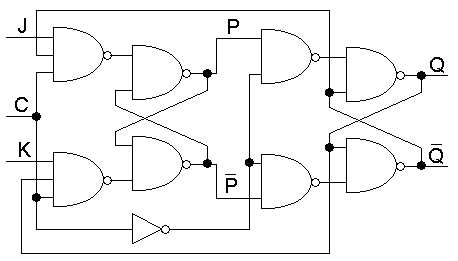}}}%
\FRAME{ftbpFUX}{1.3162in}{0.9184in}{0pt}{\Qcb{The symbol of the JK flip-flop}%
}{}{f9b.jpg}{\raisebox{-0.9184in}{\includegraphics[height=0.9184in]{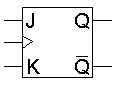}}}%
\begin{equation*}
P(t)\cdot J(t)\cdot \overline{Q(t)}\cup \overline{P(t)}\cdot K(t)\cdot
Q(t)\cup
\end{equation*}%
\begin{equation*}
\cup (\overline{P(t-0)}\cdot \overline{P(t)}\cup P(t-0)\cdot P(t))\cdot (%
\overline{J(t)}\cdot \overline{K(t)}\cup \overline{J(t)}\cdot \overline{Q(t)}%
\cup \overline{K(t)}\cdot Q(t))
\end{equation*}%
only one of $P(t)\cdot J(t)\cdot \overline{Q(t)}$, $\overline{P(t)}\cdot
K(t)\cdot Q(t)$ can be $1$, thus $P$ changes value at most once and this was
not the case at the edge triggered RS flip-flop. Let's make now in the
equations of the D flip-flop $D(t)=J(t)\cdot \overline{Q(t)}\cup \overline{%
K(t)}\cdot Q(t)$. We get%
\begin{equation*}
C(t)\cdot (\overline{Q(t-0)}\cdot \overline{Q(t)}\cup Q(t-0)\cdot Q(t))\cdot
(P(t)\cdot J(t)\cdot \overline{Q(t)}\cup \overline{P(t)}\cdot K(t)\cdot
Q(t)\cup
\end{equation*}%
\begin{equation}
\cup \overline{P(t)}\cdot \overline{J(t)}\cdot \overline{Q(t)}\cup P(t)\cdot 
\overline{K(t)}\cdot Q(t))\cup
\end{equation}%
\begin{equation*}
\cup \overline{C(t)}\cdot (\overline{Q(t)}\cdot \overline{P(t-0)}\cdot 
\overline{P(t)}\cup Q(t)\cdot P(t-0)\cdot P(t))=1
\end{equation*}%
Equations (8.2) and (8.3) have similarities and sometimes the equation of
the JK flip-flop is considered to be (8.3).

\section{T flip-flop}

The next equivalent statements:\FRAME{ftbpFUX}{4.9709in}{2.6351in}{0pt}{\Qcb{%
The T flip-flop circuit}}{}{t_flip_flop1.jpg}{\raisebox{-2.6351in}{\includegraphics[height=2.6351in]{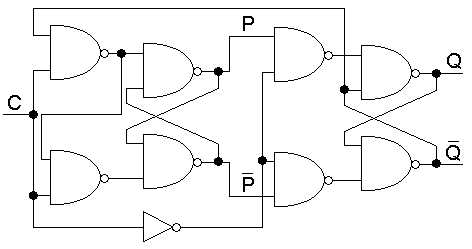}}}%
\FRAME{ftbpFUX}{1.2531in}{0.9072in}{0pt}{\Qcb{The symbol of the T flip-flop}%
}{}{f10b.jpg}{\raisebox{-0.9072in}{\includegraphics[height=0.9072in]{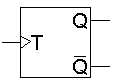}}}%
\begin{equation}
\left\{ 
\begin{array}{c}
\overline{P(t-0)}\cdot P(t)=\overline{P(t-0)}\cdot \overline{Q(t)}\cdot C(t)
\\ 
P(t-0)\cdot \overline{P(t)}=P(t-0)\cdot Q(t)\cdot C(t) \\ 
\overline{Q(t-0)}\cdot Q(t)=\overline{Q(t-0)}\cdot P(t)\cdot \overline{C(t)}
\\ 
Q(t-0)\cdot \overline{Q(t)}=Q(t-0)\cdot \overline{P(t)}\cdot \overline{C(t)}%
\end{array}%
\right.
\end{equation}%
respectively%
\begin{equation}
C(t)\cdot (\overline{Q(t-0)}\cdot \overline{Q(t)}\cdot P(t)\cup Q(t-0)\cdot
Q(t)\cdot \overline{P(t)})\cup
\end{equation}%
\begin{equation*}
\cup \overline{C(t)}\cdot (\overline{Q(t)}\cdot \overline{P(t-0)}\cdot 
\overline{P(t)}\cup Q(t)\cdot P(t-0)\cdot P(t))=1
\end{equation*}%
are called the equations of the T flip-flop. $C,P,Q$ are signals: the clock
input, the next state and the state. The conditions of admissibility of the
inputs are fulfilled for the first two and for the last two equations from
(9.1) (the master and the slave latch). At each falling edge $C(t-0)\cdot 
\overline{C(t)}=1$ of the clock input, the state $Q$ of the T flip-flop
toggles to its complementary value, otherwise it is constant. The equations
of the T flip-flop represent the next special cases: in the equations of the
edge triggered RS flip-flop, $S(t)=\overline{Q(t)},R(t)=Q(t)$; in the
equations of the D flip-flop $D(t)=\overline{Q(t)}$; in the equations of the
JK flip-flop (any of (9.2), (9.3)) \ $J(t)=1,K(t)=1$.

\section{Conclusions}

Digital electrical engineering is a non-formalized theory, where the latches
are fundamental circuits. In our work we have given the general form of the
equations that model the ideal latches, together with the theorem that
characterizes the existence and the uniqueness of the solution. Furthermore,
we have shown the manner in which this system of equations is particularized
in the case of the most well known latches and flip-flops.

The bibliography dedicated to the latches is rich and descriptive
(non-formalized). We have indicated at the references a source of
inspiration that has created some order in our thoughts.

A possibility of continuing the present ideas is that of considering models
of inertial latches, for example we can replace (1.1) with 
\begin{equation*}
\left\{ 
\begin{array}{c}
\overline{x(t-0)}\cdot x(t)=\overline{x(t-0)}\cdot \underset{\xi \in \lbrack
t-d,t)}{\bigcap }u(\xi ) \\ 
x(t-0)\cdot \overline{x(t)}=x(t-0)\cdot \underset{\xi \in \lbrack t-d,t)}{%
\bigcap }v(\xi ) \\ 
\underset{\xi \in \lbrack t-d,t)}{\bigcap }u(\xi )\cdot \underset{\xi \in
\lbrack t-d,t)}{\bigcap }v(\xi )=0%
\end{array}%
\right.
\end{equation*}%
where $d>0$. We remark that this model replaces $u$ ($v$) with $\underset{%
\xi \in \lbrack t-d,t)}{\bigcap }u(\xi )$ (with $\underset{\xi \in \lbrack
t-d,t)}{\bigcap }v(\xi )$) meaning that the $1$ value of $u$ (of $v$)
continues to produce the switch of $x$ from $0$ to $1$ (from $1$ to $0$),
but this happens only if it is persistent, i.e. if it lasts at least $d$
time units.

\end{document}